\documentstyle[12pt,twoside,epsfig]{article}
\begin{document}

\title{\vspace*{-2cm}
In-medium modification of the \\ $\Delta(1232)$ resonance at SIS energies
\footnote{Talk presented at the 15th Winter Workshop on Nuclear Dynamics
Park City, Utah, January 9-16, 1999. To be published in the Conference
Proceedings by Kluwer Academic Press.}}

\author{D. Pelte \\
Physics Institute, University of Heidelberg \\
Philosophenweg 12 \\
D-69120 Heidelberg, Germany \\
email: Dietrich.Pelte@mpi-hd.mpg.de}
\date{{\small 11. February 1999}}

\maketitle

\abstract{
From the pion production rates $n_{\pi} / A_{part}$ and the pion charge ratios
$R_{\pi} = n_{\pi^-} / n_{\pi^+}$ and $S_{\pi} = n_{\pi^0} / (n_{\pi^-} +
n_{\pi^+})$, measured in p + a and A + A reactions, it is concluded that the
mass of the $\Delta(1232)$ resonance is reduced in dense nuclear matter. This
conclusion is supported by the $\Delta(1232)$ mass distribution extracted from
correlated (p,$\pi^{\pm}$) pairs and the $\pi^{\pm}$ transverse momentum
distributions.}

\section{Introduction}
The experimental verification that the properties of hadrons are modified in
the dense nuclear medium produced by relativistic nucleus-nucleus collisions
is of great interest since the partial restoration of chiral symmetry
would lead to such modifications, in particular it predicts the reduction of
the K$^-$ mass \cite{cas99}. In this contribution I will survey the
experimental information with regard to pions produced in p + A and A + A
reactions at SIS energies. I will show that these data can be interpreted
as the result of a $\Delta(1232)$ mass reduction. This interpretation is
based on two simple models, the thermal and the isobar models, and the
conclusion is confirmed by the direct measurement of the $\Delta(1232)$ mass
distribution.

\section{The Thermal Model}
In order to obtain the charge states of hadrons from the thermal model one has
to introduce the isospin chemical potential $\mu_I$, besides the baryon chemical
potential $\mu_B$ and the temperature $T$. Defining $x = e^{-\mu_I / T}$ one
obtains for the charges states of nucleons and pions:
\begin{equation}
n_p = \frac{n_N}{2} x^{+1} \;\; , \;\; n_n = \frac{n_N}{2} x^{-1}
\end{equation}
\begin{equation}
n_{\pi^+} = \frac{n_{\pi}}{3} x^{+2} \;\; , \;\;
n_{\pi^0} = \frac{n_{\pi}}{3} \;\; , \;\;
n_{\pi^-} = \frac{n_{\pi}}{3} x^{-2}
\end{equation}
\begin{equation}
n_{\Delta^{++}} = \frac{n_{\Delta}}{4} x^{+3} ,
n_{\Delta^{+}} = \frac{n_{\Delta}}{4} x^{+1} ,
n_{\Delta^{0}} = \frac{n_{\Delta}}{4} x^{-1} ,
n_{\Delta^{-}} = \frac{n_{\Delta}}{4} x^{-3} ,
\end{equation}
where $n_N$, $n_{\pi}$, and $n_{\Delta}$ are determined by $\mu_B$ and $T$.
Since charge and baryon number are conserved the following relations between
$x$, the pion production rate $n_{\pi} / A_{part}$, and the N/Z ratio
$\zeta_{part}$ of the participants should hold:
\begin{equation}
\label{eq4}
1 = \frac{\zeta_{part}+1}{2} x + \frac{n_{\pi}}{A_{part}}
\frac{\zeta_{part}+1}{3} \left( x^2 - x^{-2} \right)
\end{equation}
\begin{equation}
\label{eq5}
1 = \frac{\zeta_{part}+1}{2} x + \frac{n_{\pi}}{A_{part}}
\frac{\zeta_{part}+1}{4} \left( 2 x^3 - x - x^{-3} \right) \;\; .
\end{equation}
The first equation is valid in case the participant region only contains
nucleons and pions, the second if it contains only nucleons and $\Delta(1232)$
resonances, where the $\Delta(1232)$ after freeze-out decay into pions. The
importance of these equations is due to the fact that they only depend on
known ($\zeta_{part}$) and measured ($n_{\pi} / A_{part}$) quantities, and
that $x$ can be deduced from the measured pion charge ratios $R_{\pi} =
n_{\pi^-} / n_{\pi^+}$. Note that $n_{\pi} / A_{part}$ requires to know the
production rates $n_{\pi^0} / A_{part}$ of neutral pions, these rates have
been measured for a number of A + A reactions by the TAPS collaboration
\cite{met98}, but they are unknown for the p + A reactions
considered here. In this latter case I have assumed $n_{\pi^0} = n_{\pi^-} +
n_{\pi^+}$. This assumption has only little consequences for the present
discussion, but its experimental confirmation would add support to the
interpretation of the existing data advocated in this report.
\begin{table}
\hspace*{1.cm}\parbox{14.cm}{
\caption{Participant temperatures $T$ extracted from different observables
assuming a participant density $\rho = 0.3 \rho_0$. The numbers in brackets
give the incident energy in AGeV.}
}
\begin{center}
\begin{tabular}{|c|c|c|c|c|l}
\hline
reaction \vrule height 14pt width 0pt depth 8pt
& $T(R_{\pi})$ & $T(n_{\pi})$ & $T(n_{\Delta})$ & $T(slope)$ \\
\hline
Au+Au(1.06) \vrule height 14pt width 0pt depth 8pt
& $56\pm5$ & $55\pm2$ & $66\pm5$ & $81\pm24$ \\
\hline
Ni+Ni(1.06) \vrule height 14pt width 0pt depth 8pt
&  $<50$   & $62\pm2$ & $75\pm3$ & $79\pm10$ \\
\hline
Ni+Ni(1.45) \vrule height 14pt width 0pt depth 8pt
&  $<50$   & $68\pm2$ & $84\pm4$ & $84\pm10$ \\
\hline
Ni+Ni(1.93) \vrule height 14pt width 0pt depth 8pt
&  $<55$   & $73\pm2$ & $93\pm3$ & $92\pm12$ \\
\hline
\end{tabular}
\end{center}
\end{table}

The data have been measured for Ni + Ni \cite{pel97a}, Au + Au \cite{pel97b},
and p + C/Nb/Pb \cite{lem88} reactions in the energy range between 1 to 2 AGeV
and at several impact parameters $b$, I have chosen central collisions by
extrapolating to $b \rightarrow 0$. The extrapolated results, which include
the production rates of the $\Delta(1232)$ resonance \cite{hon97} \cite{esk98},
are in conflict with the equations \ref{eq4} \ref{eq5}, if one
assumes for each reaction a unique temperature and baryon chemical potential.
In order to be consistent with charge and baryon number conservation the
different observables would require in general different temperatures which
are shown in table 1. In this table I have also included the temperature
values which were deduced from the slopes of the energy spectra of various
particles emitted from the participant region \cite{eos95} \cite{hon98}.
One finds that the slope temperatures are within errors consistent with 
$T(n_{\Delta})$, but that in general
\begin{equation}
T(R_{\pi}) < T(n_{\pi}) < T(n_{\Delta})  .
\end{equation}
In case of p + A reactions the data would not allow to obtain any
meaningful solution to the equations \ref{eq4} \ref{eq5}. This implies that
within this
very simple framework the thermal model is unable to explain the measured
data, and the problem can be traced in case of the A + A reactions to the
measured number of pions which is too small to accommodate the measured number
of $\Delta(1232)$ resonances and the calculated number of thermally produced
pions. In addition the $R_{\pi}$ ratios suggest that their values largely
depend on the type of charge exchange reactions that occur between the pions
and the nucleons of the participant region.
\begin{figure}[t]
\epsfxsize=13.cm
\epsffile[-100 30 570 350]{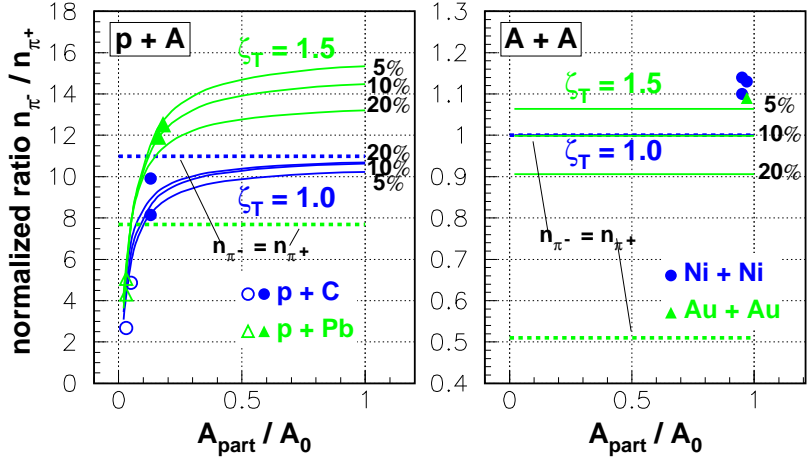}
\hspace*{1.cm}\parbox{14.cm}{
\caption{Left panel: The $A_{part} / A_0$ dependence of the normalized
$\pi^-$ to $\pi^+$ ratio $R_{\pi}^{(0)}$ of p + A reactions for two different
values $\zeta_T$ of the target neutron to proton ratio. The percentage numbers
give the probability to excite the $\Delta(1232)$ resonance in the first
step, the open symbols correspond to peripheral, the full symbols to
central data from p + $^{12}$C (black) or p + $^{208}$Pb (grey) reactions. In
case of identical symbols the larger one corresponds to the higher incident
energy.  Right panel: The same for symmetric A + A reactions as specified in
table 1.}
}
\end{figure}

\section{The Isobar Model}
The isobar model allows to calculate the charge ratios of pions and nucleons
with the assumption that pions are exclusively produced by the decay of
baryon resonances $B$, which were excited in a first step by the reaction
$N + N \rightarrow N + B$. Isospin conservation yields for this first step
in the cases in which $B$ stands for the $\Delta$ resonance
\begin{eqnarray}
\label{eq6}
& & n_{\pi^-} : n_{\pi^0} : n_{\pi^+} = \\
& & (\zeta_P + \zeta_T + 10 \zeta_P \zeta_T) :
(2 + 4 \zeta_P + 4 \zeta_T + 2 \zeta_P \zeta_T) :
(10 + \zeta_P + \zeta_T) , \nonumber
\end{eqnarray}
where $\zeta_P$, $\zeta_T$ are the N/Z ratios of the projectile respectively
the target. These relative numbers are modified when the emitted pions are
reabsorbed in participant matter and form new $\Delta$ resonances which
subsequently decay by pion emission. The final pion numbers depend on the
number $n_{loop}$ of such $\pi N \Delta$ loops since two successive loops
$i$ and $i+1$ $(i < n_{loop})$ are coupled by the equations:
\begin{eqnarray}
\label{eq7}
n_{\pi^+}^{(i+1)} = \left( n_p^{(i)} ( n_{\pi^+}^{(i)} + \frac{1}{3}
n_{\pi^0}^{(i)} ) + n_n^{(i)} \frac{1}{3} n_{\pi^+}^{(i)} \right) / A_{part}
\nonumber\\
n_{\pi^0}^{(i+1)} = \left( n_p^{(i)} ( \frac{2}{3} n_{\pi^0}^{(i)} + \frac{2}{3}
n_{\pi^-}^{(i)} ) + n_n^{(i)} ( \frac{2}{3} n_{\pi^+}^{(i)} + \frac{2}{3}
n_{\pi^0}^{(i)} ) \right) / A_{part} \\
n_{\pi^-}^{(i+1)} = \left( n_p^{(i)} \frac{1}{3} n_{\pi^-}^{(i)} + n_n^{(i)}
( \frac{1}{3} n_{\pi^0}^{(i)} + n_{\pi^-}^{(i)} ) \right) / A_{part} .\nonumber
\end{eqnarray}
The $n_{\pi}$ numbers do not change anymore once $n_{loop} > 10$ in the
reactions considered here. For the $R_{\pi}$ ratios these saturation values
are shown in Fig.1, and for the ratios $S_{\pi} = n_{\pi^0} / (n_{\pi^-} +
n_{\pi^+})$ in Fig.2. Note that I have normalized the $R_{\pi}$ and
$S_{\pi}$ values by the predictions derived from equation \ref{eq6}, i.e.
$R_{\pi}^{(0)} = S_{\pi}^{(0)} = 1$ would imply that the value of $n_{loop}$
is sufficiently small not to modify the ratios of the first step. The
saturation values of $R_{\pi}^{(0)}$ and $S_{\pi}^{(0)}$ depend in case of
p + A reactions on the size $A_{part}$ of the participant, but not in case
of A + A reactions. The dependence on the excitation probability of the
$\Delta$ resonance, listed in Fig.1 and 2 by the percentages, is
only weak. The comparison with the data shown by symbols indicates that to
explain the p + A results one has to assume the presence of sufficiently many
$\pi N \Delta$ loops, but that in the case of A + A reactions the number of
such loops is much smaller since for both ratios one observes $R_{\pi}^{(0)}
\approx S_{\pi}^{(0)} \approx 1$.
\begin{figure}[t]
\epsfxsize=13.cm
\epsffile[-100 30 570 350]{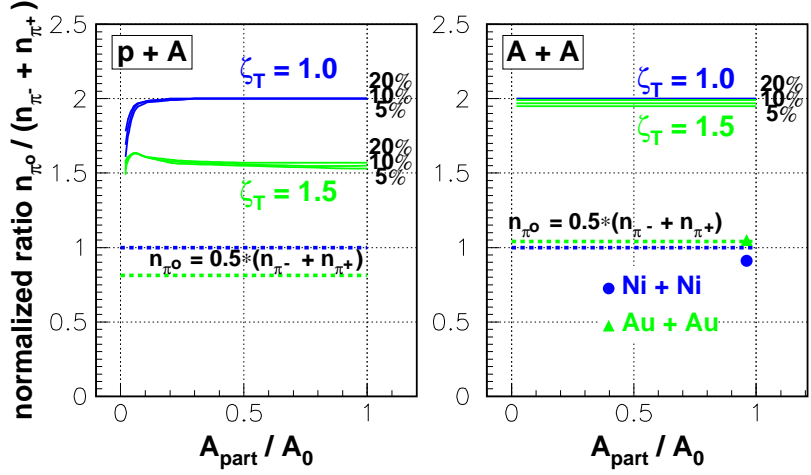}
\hspace*{1.cm}\parbox{14.cm}{
\caption{As in Fig.4 but for the normalized neutral to charged
pion ratio $S_{\pi}^{(0)}$. Experimental data (symbols) only exist for A + A
reactions.}
}
\end{figure}

There exist 2 possible explanations for this difference between the behavior
of p + A and A + A reactions:
\begin{itemize}
\item{The decay probability of $\Delta(1232)$ resonances decreases in
participant matter which reduces the number of $\pi N \Delta$ loops,}
\item{the absorption probability of $\Delta(1232)$ resonances via the
$\Delta + N \rightarrow N + N$ reaction increases in participant matter
which also reduces the number of $\pi N \Delta$ loops.}
\end{itemize}
Of these two alternatives the second explanation is favored by the experimental
observation that the pion production rate depends on the total system mass
$A_0 = A_P + A_T$. The measured $n_{\pi} / A_{part}$ values,
integrated over the impact parameter, may
be fitted for incident energies from 1 to 2 AGeV by the relation \cite{har87}
\begin{eqnarray}
\label{eq8}
\frac{ \langle n_{\pi} \rangle}{\langle A_{part} \rangle} =
a_{\pi} \cdot (E_{kin} - 0.11) ,
\end{eqnarray}
where the cm energy $E_{kin}$ is given in AGeV. The production coefficient
$a_{\pi}$ is shown in Fig.3, it displays a linear decrease with rising $A_0$.
In addition the $n_{\pi}$ rate from p + p reactions is noticeably enhanced
compared to A + A reactions \cite{gaz98}. It is evident that compared to p + p
reactions the pion production rates are suppressed in A + A reactions, and the
suppression is the stronger the heavier the system is, or the larger the
participant density is which can be attained in nucleus-nucleus collisions.
More insight into the mechanism which might cause the rise of the absorption
cross section is gained by studying the $\Delta(1232)$ mass distribution in
participant matter.
\begin{figure}
\epsfxsize=10.cm
\epsffile[0 80 600 490]{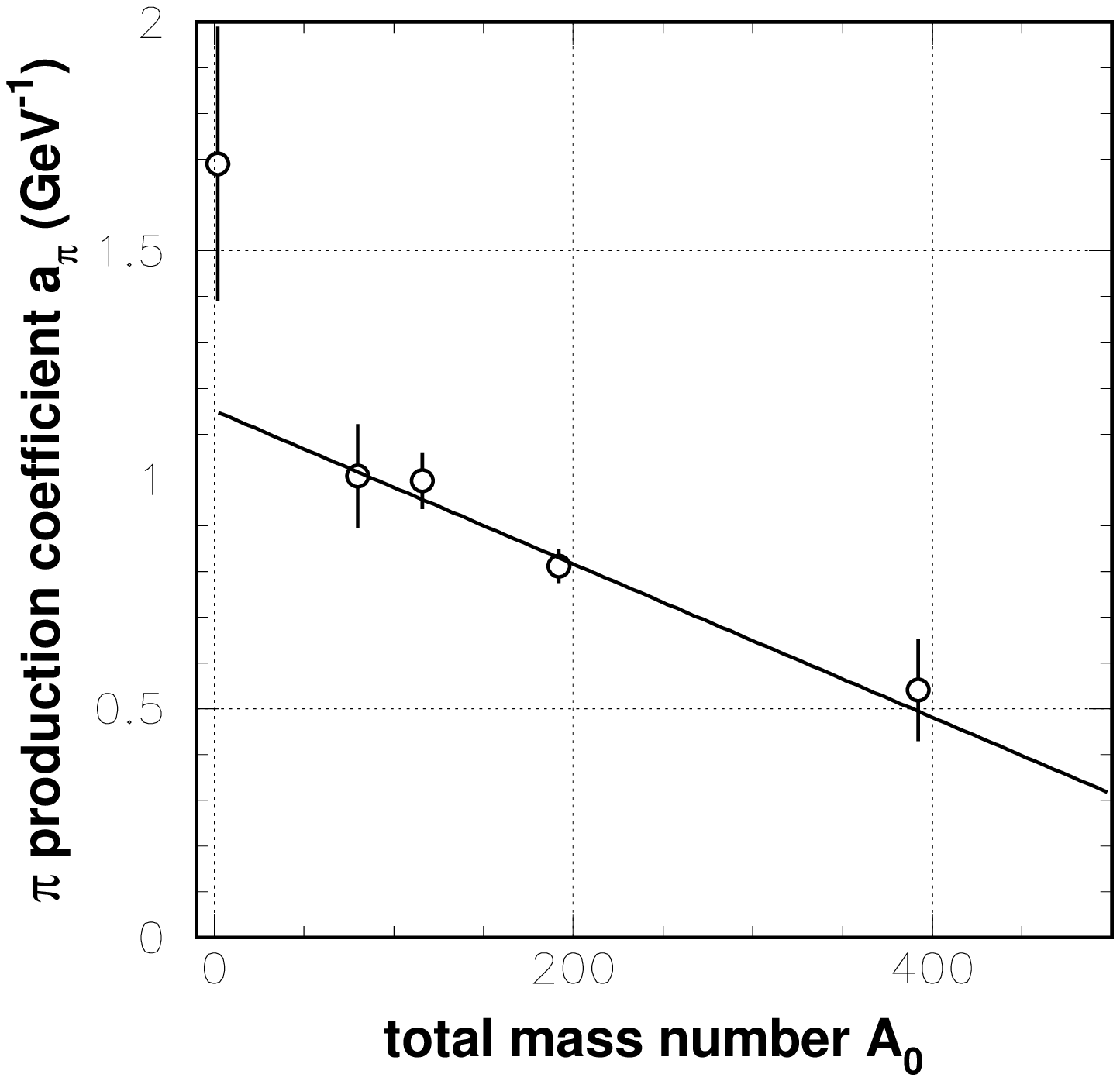}
\vspace*{-3.8cm}
\hspace*{9.cm}\parbox{6.cm}{
\caption{The measured pion production coefficients $a_{\pi}$ as function
of the system mass $A_0$. The line is a linear fit through all points except
the one at $A_0 = 2$.}
}
\vspace*{0.5cm}
\end{figure}

\section{The $\Delta(1232)$ Mass Distribution}
The mass distribution of the $\Delta(1232)$ resonance in participant matter
can be reconstructed from the transverse momentum spectra $dn / dp_t$ of pions,
or from the invariant mass of (p,$\pi$) pairs. The former technique was
applied in A + A reactions \cite{esk98}, the latter also in p + A reactions
\cite{trz91}. I display the results for central Ni + Ni and Au + Au reactions
in Fig.4 lower panels, where the dark points correspond to the $dn / dp_t$
technique and the stars to the (p,$\pi$) technique.
For the p + C reaction the equivalent data are displayed in Fig.4 upper panel.
In general the mass distributions are shifted from the
mass distribution of the free $\Delta(1232)$ resonance (dashed curves) towards
smaller masses. In case of the p + C reaction this shift can be explained
in the framework of the thermal model by
the finite temperature $T = 65$ MeV of the participants, the deviations from
the expected mass distribution at higher masses are most likely due to
first-chance N + N collisions. Contrary to these findings, in Ni + Ni and Au +
Au reactions the observed mass shifts are only partly reproduced by the
participant temperatures $T(n_{\pi})$ of table 1 (dotted curves). To obtain
a fit of the experimental mass distributions additional shifts of $\approx
-100$ MeV/c$^2$ in case of Au + Au, and $\approx -50$ MeV/c$^2$ in case of Ni
+ Ni are necessary. The data do not allow to obtain a more quantitative result,
in particular they do not allow to disentangle the medium effects onto the mass
and the width of the distribution. The size of the in-medium modification
depends on the system mass, and the result for Ni + Ni is in fair agreement
with similar data for the Ni + Cu reaction published in \cite{hjo97}.
\begin{figure}
\epsfxsize=7.8cm
\epsffile[-30 60 500 520]{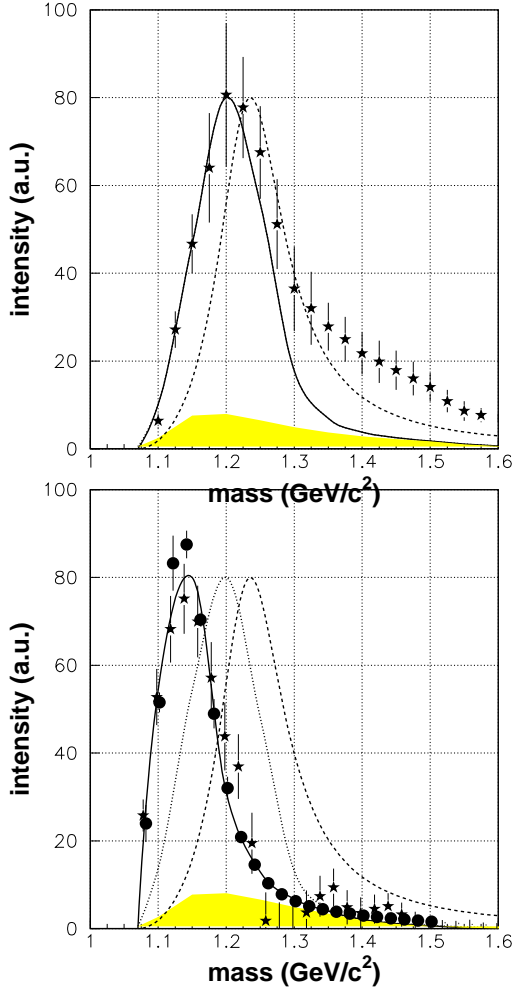}
\vspace*{-6.8cm}
\hspace*{9.cm}\parbox{6.cm}{
\caption{The invariant mass spectrum of the $\Delta(1232)$ resonance
excited in the p + C reaction at 1.6 GeV incident energy (top), the Au + Au
reaction at 1.06 AGeV (bottom left), and Ni + Ni reaction at 1.93 AGeV (bottom
right). The shaded areas correspond to the contributions from higher baryon
resonances. For the symbols and different curves see text.}
}
\vspace*{-0.5cm}
\end{figure}
\begin{figure}
\begin{minipage}{7.8cm}
\epsfxsize=7.8cm
\epsffile[-30 60 500 520]{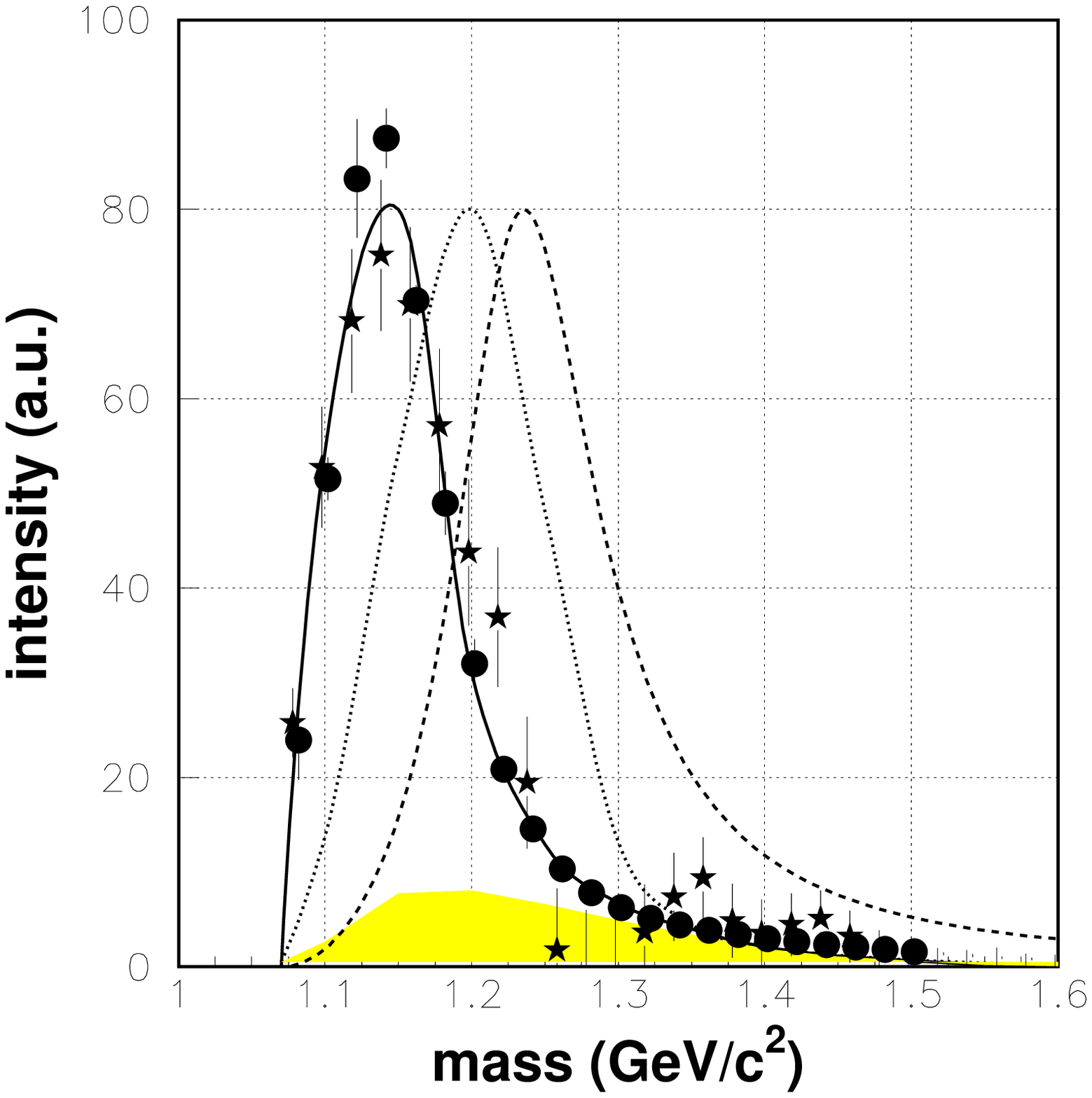}
\end{minipage}
\hspace*{-0.8cm}
\begin{minipage}{7.8cm}
\epsfxsize=7.8cm
\epsffile[-30 60 500 520]{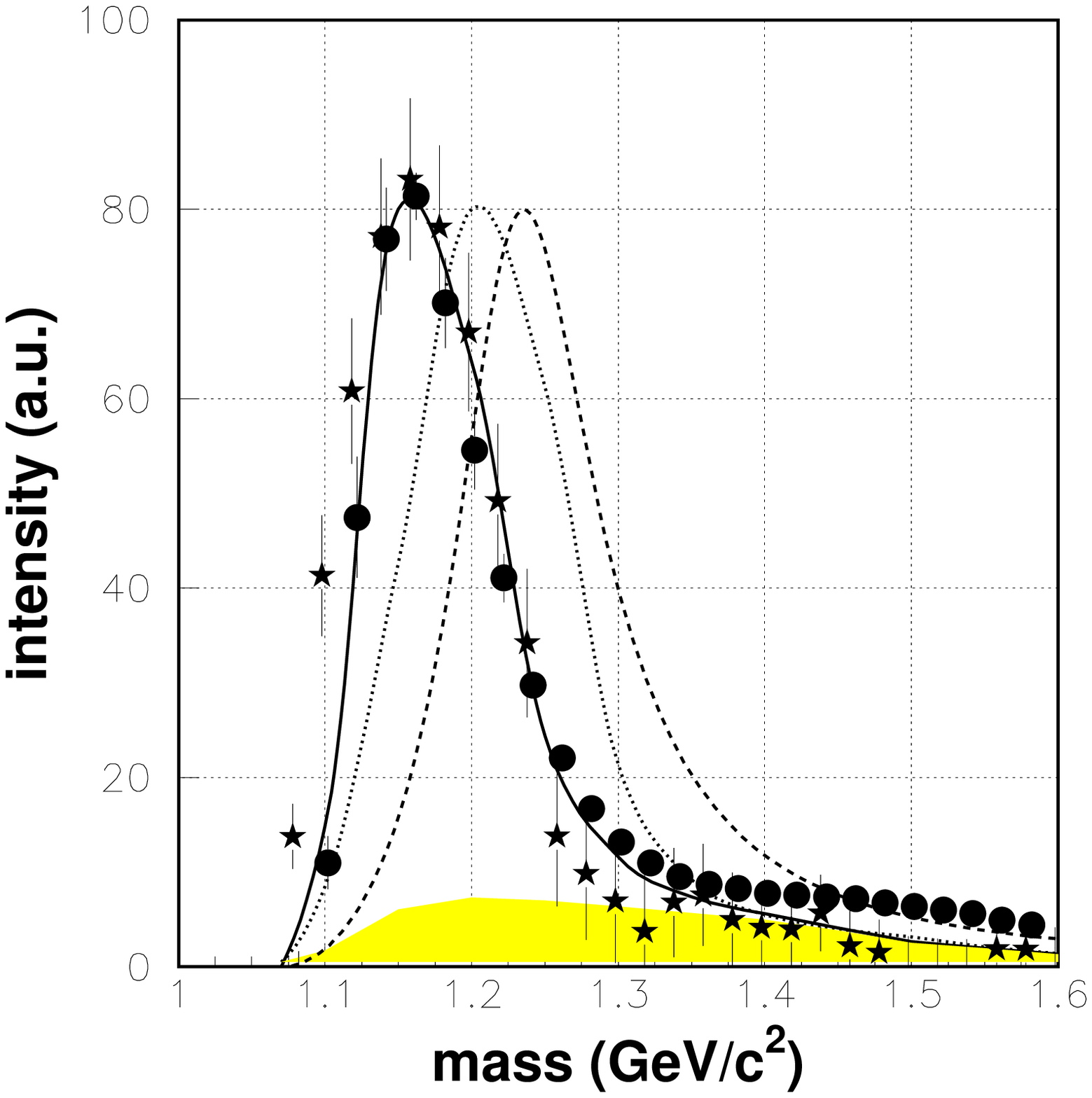}
\end{minipage}
\end{figure}

\section{Discussion}
The analyses of the pion data from p + A and A + A reactions at SIS energies
have yielded convincing evidence that \\
\hspace*{2.cm}1. the $\Delta(1232)$ absorption increases, \\
\hspace*{2.cm}2. the $\Delta(1232)$ average mass is reduced \\
as function of the average participant density which increases with the system
mass. These two phenomena are indeed related by the ''extended principle of
detailed balance`` via
\begin{equation}
\label{eq9}
\sigma_{n \Delta^{++} \rightarrow p p} = \frac{1}{4} \frac{p_N^2}{p_{\Delta}^2}
\sigma_{p p \rightarrow n \Delta^{++}} \frac{1}{\int_{(m_N + m_{\pi})^2}
^{(\sqrt{s} - m_N)^2} f_{\Delta}(m^2) \, dm^2} \;\; ,
\end{equation}
where $\sigma_{n \Delta^{++} \rightarrow p p}$ is the partial absorption cross
section for the $\Delta(1232)$ resonance in the reaction $\Delta + N
\rightarrow N + N$, and $f_{\Delta}(m^2)$ is the in-medium mass distribution
of the $\Delta(1232)$ resonance. The equation \ref{eq9} has been studied in
\cite{dan91} and \cite{wol93}, the differences in these studies have to be
resolved before a detailed comparison with the experimental results can be
accomplished. It is of interest that in \cite{hjo97} the mass shift was found
to become weaker with rising impact parameter. In terms of the presently
advocated interpretation this implies a reduction of the pion absorption.
Experimentally the pion production rate $n_{\pi} / A_{part}$ depends on
$A_{part}$, however the dependence is different for $\pi^-$ and $\pi^+$
\cite{pel97a} \cite{pel97b}. Therefore Coulomb effects have to play an
important role in the pion dynamics.

\section*{Acknowledgments}
The author wishes to thank the FOPI and the TAPS collaborations which supplied
most of the data used in this study. Partial funding was provided by contracts
06 HD 525 I and HD Pel K.

\end{document}